\documentclass[pra,aps, twocolumn,showpacs]{revtex4}
\usepackage{amsmath}
\usepackage{amssymb}
\usepackage{epsfig}

\begin{document}

\title{Quantification of Complementarity in Multi-Qubit Systems}
\author{Xinhua \surname{Peng}$^{1,2,}$ \thanks{%
1}}
\author{Xiwen \surname{Zhu}$^{2}$ }
\author{Dieter \surname{Suter}$^{1}$ }
\author{Jiangfeng \surname{Du}$^{3}$ }
\author{Maili \surname{Liu}$^{2}$ }
\author{Kelin \surname{Gao}$^{2}$ }
\affiliation{$^{1}$Fachbereich Physik, Universit\"{a}t Dortmund, 44221 Dortmund, Germany}
\affiliation{$^{2}$Wuhan Institute of Physics and Mathematics, Chinese Academy of
Sciences, Wuhan 430071, P. R. China}
\affiliation{$^{3}$Hefei National Laboratory for Physical Sciences at Microscale and
Department of Modern Physics, University of Science and Technology of China,
Hefei, Anhui 230026, P.R. China}
\date{\today}

\begin{abstract}
Complementarity was originally introduced as a qualitative concept for the
discussion of properties of quantum mechanical objects that are classically
incompatible. More recently, complementarity has become a \emph{quantitative}
relation between classically incompatible properties, such as visibility of
interference fringes and "which-way" information, but also between purely
quantum mechanical properties, such as measures of entanglement. We discuss
different complementarity relations for systems of 2-, 3-, or \textit{n}
qubits. Using nuclear magnetic resonance techniques, we have experimentally
verified some of these complementarity relations in a two-qubit system.
\end{abstract}

\pacs{03.65.Ta, 03.65.Ud, 76.60.-k}
\maketitle

\section{Introduction}

Complementarity is one of the most characteristic properties of quantum
mechanics, which distinguishes the quantum world from the classical one. In
1927, Bohr\cite{Bohr} first reviewed this subject, observing that the wave-
and particle-like behaviors of a quantum mechanical object are mutually
exclusive in a single experiment, and referred to this as complementarity.
Probably the most popular representation of Bohr complementarity is the
`wave-particle duality'\cite{Feynman,Scully}, which is closely related to
the long-standing debate over the nature of light \cite{Simonyi}. This type
of complementarity is often illustrated by means of two-way interferometers:
A classical particle can take only one path, while a classical wave can pass
through both paths and therefore display interference fringes when the two
partial waves are recombined. Depending on their state, quantum mechanical
systems (quantons) can behave like particles (go along a single path), like
waves (show interference), or remain in between these extreme cases by
exhibiting particle- as well as wave-like behavior. This can be quantified
by the predictability $P$, which specifies the probability that the system
will go along a specific path, and the visibility $V$ of the interference
fringes after recombination of the two partial waves, which quantifies the
wavelike behavior. A quantitative expression for the complementarity is the
inequality \cite{Wootters,Bartell,Greenberger,Mandel,JaegerSV,Englert} 
\begin{equation}
P^{2}+V^{2}\leq 1,  \label{e.vp}
\end{equation}%
which states that the more particle-like a system behaves, the less
pronounced the wave-like behavior becomes.

In composite systems, consisting of two (or more) quantons, it is possible
to optimize the \textquotedblleft which-way" information of one particle:
one first performs an ideal projective measurement on the second particle.
By an appropriate choice of the measurement observable, one can then
maximize the predictability for the first partial system. This optimized
property, which is called distinguishability $D$, obeys a similar inequality 
\cite{Wootters,Bartell,Greenberger,Mandel,JaegerSV,Englert}: 
\begin{equation}
D^{2}+V^{2}\leq 1.  \label{e.dv}
\end{equation}%
For pure states, the limiting equality holds, 
\begin{equation}
D^{2}+V^{2}=1 ,  \label{e.dvp}
\end{equation}%
while the inequality holds for mixed states. This issue has been
experimentally investigated in the context of interferometric experiments,
using a wide range of physical objects including photons\cite{Taylor},
electrons\cite{Mollenstedt}, neutrons\cite{Zeilinger}, atoms\cite{Carnal}
and nuclear spins in a bulk ensemble with nuclear magnetic resonance (NMR)
techniques \cite{Zhu,Peng}.

In systems of strongly correlated pairs of particles, it is often useful to
consider particle pairs as composite particles with an independent identity.
Such composite particles that consist of identical particles include pairs
of electrons (Cooper pairs) and photon pairs\cite{Ghosh}. Many interesting
phenomena, such as superconductivity, are much easier to understand in terms
of the composite particles than in terms of the individual particles.
Suitable experiments, such as two-photon interference \cite{Ghosh,All} can
measure properties of the composite particles. These experiments made it
possible to quantify the ``compositeness" of a two-particle state. Extreme
cases are product states, which show no signal in two-particle interference
experiments, while maximally entangled states maximize the two-particle
visibility but show vanishing visibility in experiments testing the
interference of individual particles \cite{HorneZeil}. Between these
extremes lies a continuum of states for which the complementarity relation 
\begin{equation}
V_{_{k}}^{2}+V_{_{12}}^{2}\leq 1,(k=1,2)  \label{e.vv}
\end{equation}
holds, which is valid for bipartite pure states\cite{JaegerSV,JaegerHS}.
Here, $V_k$ is the single-particle visibility for particle $k$, while $%
V_{12} $ represents the two-particle visibility. This intermediate regime of
the complementarity relation of one- and two-photon interference has only
recently been experimentally demonstrated in a Young's double-slit
experiment by Abouraddy et al.\cite{Abouraddy}.

From a quantum information theoretic point of view, composite quantum
systems involve inevitably the concept of entanglement, which is a uniquely
quantum resource with no classical counterpart. Does entanglement constitute
a physical feature of quantum systems that can be incorporated into the
principle of complementarity? Some authors have explored this question and
obtained some important results, such as the complementarities between
distinguishability and entanglement\cite{OHHHH}, between coherence and
entanglement\cite{SAST} and between local and nonlocal information\cite{BH}
etc. Additionally, some complementarity relations in \textit{n}-qubit pure
systems are also observed such as the relationships between multipartite
entanglement and mixedness for special classes of \textit{n}-qubit systems%
\cite{JSST}, and between the single particle properties and the \textit{n}
bipartite entanglements in an arbitrary pure state of \textit{n} qubits\cite%
{Tessier}.

More recently, Jakob and Bergou\cite{JB} derived a generalized duality
relation between bipartite and single partite properties for an arbitrary
pure state of two qubits, which in some sense accounts for many previous
results. They showed that an arbitrary normalized pure state $\left\vert
\Theta \right\rangle $ of a two-qubit system satisfies the expression\cite%
{JB} : 
\begin{equation}
C^{2}+V_{k}^{2}+P_{k}^{2}=1.  \label{e.cvp}
\end{equation}%
Here the concurrence $C$\cite{Wootters1998,Wootters2001} is defined by 
\begin{equation}
C\left( \left\vert \Theta \right\rangle \right) \equiv \left\vert
\left\langle \Theta \right\vert \left( \sigma _{y}^{\left( 1\right) }\otimes
\sigma _{y}^{\left( 2\right) }\right) \left\vert \Theta ^{\ast
}\right\rangle \right\vert  \label{e.conc_p}
\end{equation}%
as a measure of entanglement. $\sigma _{y}^{\left( k\right) }$ is the y
component of the Pauli operator on qubit $k$ and $\left\vert \Theta ^{\ast
}\right\rangle $ is the complex conjugate of $\left\vert \Theta
\right\rangle $. The concurrence is a bipartite quantity, which quantifies
quantum \emph{nonlocal} correlations of the system and is taken as a measure
of the \emph{bipartite} character of the composite system. The complement 
\begin{equation}
S_{k}^{2}=V_{k}^{2}+P_{k}^{2}  \label{e.Sk}
\end{equation}%
combines the single-particle fringe visibility $V_{k}$ and the
predictability $P_{k}$. This quantity is invariant under local unitary
transformations (though $V_{k}$ and $P_{k}$ are not), and is therefore taken
as a quantitative measure of the \emph{single-particle} character of qubit $%
k $.

Since the two-particle visibility is equal to the concurrence, $V_{12}\equiv
C$ \cite{JB}, we can rewrite Eq. (\ref{e.cvp}) as 
\begin{equation}
V_{12}^{2}+V_{k}^{2}+P_{k}^{2}=1,\quad \left( k=1,2\right) .  \label{e.Vvp}
\end{equation}%
This turns the inequality (\ref{e.vv}) into an equality and identifies the
missing quantity as the predictability $P_{k}$.

For pure bipartite systems, an equation similar to Eq. (\ref{e.dvp}) holds, $%
D_{k}^{2}+V_{k}^{2}=1$. Here, the index $k=1,2$ refers to the different
particles as the interfering objects in the bipartite system. Combining this
with Eq. (\ref{e.cvp}), we obtain 
\begin{equation}
D_{k}^{2}=P_{k}^{2}+C^{2}.  \label{D_pc}
\end{equation}%
Apparently, $D_{k}$ contains both the \textit{a priori} WW information $P_{k}
$ and the additional information encoded in the quantum correlation to an
additional quantum system which serves as the possible information storage.
This quantum correlation can be measured by the concurrence. This reveals
explicitly that quantum correlation can help to optimize the information
that can be obtained from a suitable measurement; without entanglement, the
available WW information is limited to the \textit{a priori} WW knowledge $%
P_{k}$.

For \textit{mixed states}, a weaker statement for the complementarity (\ref%
{e.cvp}) is found in the form of an inequality $C^{2}+V_{k}^{2}+P_{k}^{2}%
\leq 1$. However, there is no corresponding inequality for the two-particle
visibility $V_{12}$ in the mixed two-particle sources because it is very
difficult to get a clear and definite expression for $V_{12}$ and the direct
relation between concurrence and two-particle visibility ceases to exist for
mixed states\cite{JB}.

In this paper, we give a proof-of-principle experimental demonstration of
the complementarities (\ref{e.dvp}), (\ref{e.cvp}) and (\ref{e.Vvp}) in a
two-qubit system. In addition, we extend the complementarity relation (\ref%
{e.cvp}) to multi-qubit systems. The remainder of the paper is organized as
follows: In Sec II, we introduce NMR interferometry as a tool for measuring
visibilities and which-way information. Sec. III and Sec. IV discuss
measurements of the visibilities and the "which-way" information in pure
bipartite systems. Sec. V is an experimental investigation of the
complementarity relation for a pure bipartite system on the basis of
liquid-state NMR. For this purpose, we express the entanglement
(concurrence) in terms of directly measurable quantities: the two-particle
visibility $V_{12}$ and the distinguishability $D_{k}$. This allows us to
test two interferometric complementarities (\ref{e.Vvp}) and (\ref{e.dvp})
by specific numerical examples. In section VI we generalize the
complementarity relation (\ref{e.cvp}) to multi-qubit systems. A
quantitative complementarity relation exists between the single-particle
property and the bipartite entanglement between the particle and the
remainder of the system in pure multi-qubit systems. This allows us to
derive, for pure three-qubit system, a relation between the single-particle,
bipartite and tripartite properties, which should generalize to arbitrary
pure states of $n$ qubit systems. Finally, a brief summary with a discussion
is given in Sec. VII.\bigskip 

\section{NMR interferometry}

Complementarity relations are often discussed in terms of photons or other
particles propagating along different paths. Another, very flexible approach
is to simulate these systems in a quantum computer. In particular
liquid-state NMR has proved very successful for such investigations. Optical
interferometers can readily be simulated by NMR-interferometry \cite{Su88}.

\begin{figure}[tbh]
\begin{center}
\includegraphics[width = 0.8\columnwidth]{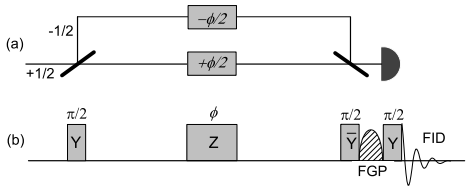}
\end{center}
\caption{Principle of NMR interferometry: (a) Path representation and (b)
Pulse sequence. }
\label{NMR_int}
\end{figure}

Figure \ref{NMR_int} shows how such an interferometric experiment can be
implemented by a sequence of radio-frequency pulses. Assuming an ideal spin $%
I=\frac{1}{2}$ particle, the Hilbert space $\mathcal{H}_{1}$ associated with
the particle is spanned by vectors $\left\vert 0\right\rangle (m = +\frac{1}{%
2})$ and $\left\vert 1\right\rangle (m = -\frac{1}{2})$. A beam splitter,
which puts the particle incoming from one port into a superposition of both
paths is realized by a radio frequency pulse that puts the spin in a
superposition of the two basis states. If the flip angle of the pulse is
taken as $\frac{\pi }{2}$, it corresponds to a \emph{symmetric} beam
splitter. A relative phase shift between the two paths, which corresponds to
a path length difference, can be realized by a rotation of the spin around
the z-axis. The second $\frac{\pi }{2}$ radio frequency pulse recombines the
two paths.

For the discussion of the complementarity of interference vs. ``which-way"
information, we consider the superposition state behind the first beam
splitter as the starting point. The action of the phase shifter and the
second beam splitter can then be summarized into a transducer.
Mathematically, this transducer maps the input state into an output state by
the transformation 
\begin{equation}
U\left( \phi \right) =e^{i\frac{\pi }{4}\sigma _{y}}e^{-i\frac{\phi }{2}%
\sigma _{z}}=\frac{1}{\sqrt{2}}\left( 
\begin{array}{cc}
e^{-i\phi /2} & e^{i\phi /2} \\ 
-e^{-i\phi /2} & e^{i\phi /2}%
\end{array}
\right) .  \label{Uk.op}
\end{equation}

In the NMR interferometer, a number of different possibilities exist for
implementing the action of the transducer. We chose the following pulse
sequence, which provides high fidelity for a large range of experimental
parameters: 
\begin{equation}
\left[ \pi \right]_{(-\pi-\phi)/2} \left[\frac{\pi}{2}\right]_{\pi/2}.
\label{pulse.Uk}
\end{equation}
Here, we have used the usual convention that $\left[\alpha \right]_{\beta}$
refers to an rf pulse with flip-angle $\alpha$ and phase $\beta$.

The resulting populations of both states in the output space vary with the
phase angle $\phi $. As shown in Fig. 1, they can be read out by first
deleting coherence with a field gradient pulse (FGP) and then converting the
population difference into observable transverse magnetization by a $\frac{%
\pi }{2}$ read-out pulse. The amplitude of the resulting FID (= the integral
of the spectrum) measures then the populations: 
\begin{equation*}
S_{NMR}\sim p(|0\rangle )-p(|1\rangle )=2p(|0\rangle )-1,
\end{equation*}%
where we have taken into account that the sum of the populations is unity.
The experimental signal can be normalized to the signal of the system in
thermal equilibrium.

\begin{figure}[tbh]
\begin{center}
\includegraphics[width = 0.8\columnwidth]{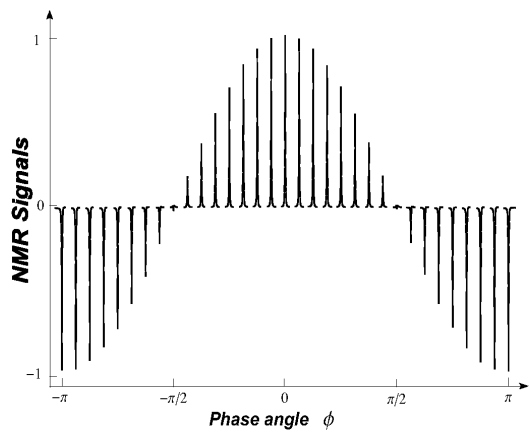}
\end{center}
\caption{NMR signals versus the phase angle $\protect\phi $. }
\label{Spectr}
\end{figure}

Figure \ref{Spectr} shows, as an example, the interference pattern for the
single proton spin in H$_{2}$O. The amplitude of the spectral line shows a
sinusoidal variation with the phase angle $\phi $, which implies the
sinusoidal variation of the population $p(|0\rangle )$ or $p(|1\rangle )$.

The visibility of the resulting interference pattern is defined as 
\begin{equation}
V=\frac{\left[ p\left( \left\vert x\right\rangle \right) \right] _{\max }-%
\left[ p\left( \left\vert x\right\rangle \right) \right] _{\min }}{\left[
p\left( \left\vert x\right\rangle \right) \right] _{\max }+\left[ p\left(
\left\vert x\right\rangle \right) \right] _{\min }}  \label{def.v}
\end{equation}%
where $x=0$ or $1$, and $p_{min}$ and $p_{max}$ are the minimal and maximal
populations (as a function of $\phi $).

Since an input state 
\begin{equation}
\rho ^{\left( i\right) }=\frac{1}{2}\left( 1+\vec{s}^{\left( i\right) }\cdot 
\vec{\sigma}\right)
\end{equation}%
with an initial Bloch vector $\vec{s}^{\left( i\right) }=\left(
s_{x}^{\left( i\right) },s_{y}^{\left( i\right) },s_{z}^{\left( i\right)
}\right) $ and Pauli spin operators $\vec{\sigma}\boldsymbol{=}\left( \sigma
_{x},\sigma _{y},\sigma _{z}\right) $ is transformed into 
\begin{equation}
\rho ^{\left( i\right) }\overset{U\left( \phi \right) }{\longrightarrow }%
\rho ^{\left( f\right) }=\frac{1}{2}\left( 1+\vec{s}^{\left( f\right) }\cdot 
\vec{\sigma}\right)
\end{equation}%
with $\vec{s}^{\left( f\right) }=\left( -s_{z}^{\left( i\right)
},s_{x}^{\left( i\right) }\sin \phi +s_{y}^{\left( i\right) }\cos \phi
,s_{x}^{\left( i\right) }\cos \phi -s_{y}^{\left( i\right) }\sin \phi
\right) $ by the transducer, we find for the visibility 
\begin{equation}
V = \sqrt{\left( s_{x}^{\left( i\right) }\right) ^{2}+\left( s_{y}^{\left(
i\right) }\right) ^{2}}
\end{equation}
and for the predictability 
\begin{equation}
P = \left\vert s_{z}^{\left( i\right) }\right\vert .
\end{equation}%
With the described experiment, it is thus straightforward to verify the
inequality (1).

\section{Visibilities in bipartite systems}

\subsection{Theory}

\begin{figure}[tbh]
\begin{center}
\includegraphics[width = 0.8\columnwidth]{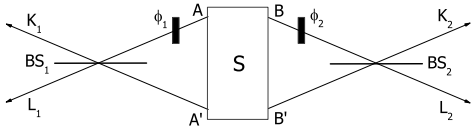}
\end{center}
\caption{Schematic two-particle interferometer using beam splitters BS$_{1}$%
, BS$_{2}$ and phase shifters $\protect\phi _{1}$, $\protect\phi _{2}$.}
\label{setup}
\end{figure}

The NMR interferometry experiment can easily be expanded to multi-qubit
systems. We start with a discussion of pure bipartite systems, where we
explore the visibility in different types of interferometric experiments,
geared towards single- and bipartite properties. Figure \ref{setup} shows
the reference setup: The source \textbf{S} emits a pair of particles 1 and
2, one of which propagates along path $A$ and/or $A^{\prime }$, through a
variable phase shifter $\phi _{1}$ impinging on an ideal beam splitter BS$%
_{1}$, and is then registered in either beam $K_{1}$ or $L_{1}$. On the
other side there is the analogous process for the other particle with paths $%
B$ and $B^{\prime }$.

Without loss the generality, we first associate states $\left\vert
A\right\rangle ,\left\vert B\right\rangle ,\left\vert K_{1}\right\rangle ,$%
and $\left\vert K_{2}\right\rangle $ in Fig. \ref{setup} with the spin-up
state $\left\vert 0\right\rangle $ and $\left\vert A^{\prime }\right\rangle $%
, $\left\vert B^{\prime }\right\rangle $, $\left\vert L_{1}\right\rangle $,
and $\left\vert L_{2}\right\rangle $ with the spin-down state $\left\vert
1\right\rangle $. A particle pair emitted from the source \textbf{S} can be
expressed as the general pure two-qubit state $\left\vert \Theta
\right\rangle $: 
\begin{equation}
\left\vert \Theta \right\rangle =\gamma _{1}\left\vert 0\right\rangle
_{1}\left\vert 0\right\rangle _{2}+\gamma _{2}\left\vert 0\right\rangle
_{1}\left\vert 1\right\rangle _{2}+\gamma _{3}\left\vert 1\right\rangle
_{1}\left\vert 0\right\rangle _{2}+\gamma _{4}\left\vert 1\right\rangle
_{1}\left\vert 1\right\rangle _{2}  \label{state}
\end{equation}%
with complex coefficients $\gamma _{i}$ that are normalized to 1.

\bigskip Assuming that the transducers consist of variable phase shifters
and \emph{symmetric} beam splitters, they can be described by the unitary
operation 
\begin{equation}
\mathcal{U}\left( \phi _{1},\phi _{2}\right) =U_{1}\left( \phi _{1}\right)
\otimes U_{2}\left( \phi _{2}\right)   \label{U.op}
\end{equation}%
where each transducer is defined according to Eq. (\ref{Uk.op}). Here the
subscripts label two different particles. Applying the transducer (\ref{U.op}%
) to the initial state (\ref{state}), we can calculate the detection
probabilities in the output channels as 
\begin{equation}
\begin{array}{l}
p\left( \left\vert x\right\rangle _{1}\right) =\frac{1}{2}+\left( -1\right)
^{x}\left\vert \gamma _{1}\gamma _{3}^{\ast }+\gamma _{2}\gamma _{4}^{\ast
}\right\vert \cos \left( \phi _{1}-\delta _{1}\right) , \\ 
p\left( \left\vert x\right\rangle _{2}\right) =\frac{1}{2}+\left( -1\right)
^{x}\left\vert \gamma _{1}\gamma _{2}^{\ast }+\gamma _{3}\gamma _{4}^{\ast
}\right\vert \cos \left( \phi _{2}-\delta _{2}\right) ,%
\end{array}
\label{e.counts}
\end{equation}%
where $x=0$ or $1$, $\gamma _{1}\gamma _{3}^{\ast }+\gamma _{2}\gamma
_{4}^{\ast }=\left\vert \gamma _{1}\gamma _{3}^{\ast }+\gamma _{2}\gamma
_{4}^{\ast }\right\vert e^{i\delta _{1}}$ and $\gamma _{1}\gamma _{2}^{\ast
}+\gamma _{3}\gamma _{4}^{\ast }=\left\vert \gamma _{1}\gamma _{2}^{\ast
}+\gamma _{3}\gamma _{4}^{\ast }\right\vert e^{i\delta _{2}}$. The single
particle count rates $p\left( \left\vert x\right\rangle _{k}\right) $ reach
their maxima and minima when the phase shifters are set to $\phi _{k}=n\pi
+\delta _{k},\left( n=0,\pm 1\right) $. From Eqs. (\ref{def.v}) and (\ref%
{e.counts}), the single particle visibilities can be obtained as 
\begin{equation}
\mathcal{V}_{1}=2\left\vert \gamma _{1}\gamma _{3}^{\ast }+\gamma _{2}\gamma
_{4}^{\ast }\right\vert \hspace{1cm}\mathcal{V}_{2}=2\left\vert \gamma
_{1}\gamma _{2}^{\ast }+\gamma _{3}\gamma _{4}^{\ast }\right\vert .
\label{e.vis}
\end{equation}

Two-particle properties can be measured by higher order correlations.
Following reference \cite{JaegerSV,JaegerHS}, we use the \textquotedblleft
corrected\textquotedblright\ two-particle fringe visibility 
\begin{equation}
\mathcal{V}_{12}=\frac{\left[ \overline{p}\left( \left\vert x\right\rangle
_{1}|y\rangle _{2}\right) \right] _{\max }-\left[ \overline{p}\left(
\left\vert x\right\rangle _{1}|y\rangle _{2}\right) \right] _{\min }}{\left[ 
\overline{p}\left( \left\vert x\right\rangle _{1}|y\rangle _{2}\right) %
\right] _{\max }+\left[ \overline{p}\left( \left\vert x\right\rangle
_{1}|y\rangle _{2}\right) \right] _{\min }}.  \label{def.v12}
\end{equation}%
where $x,y=0$ or $1$. The \textquotedblleft corrected\textquotedblright\
joint probability $\overline{p}\left( \left\vert x\right\rangle
_{1}|y\rangle _{2}\right) =p\left( \left\vert x\right\rangle _{1}|y\rangle
_{2}\right) -p\left( \left\vert x\right\rangle _{1}\right) p\left( |y\rangle
_{2}\right) +\frac{1}{4}$ are defined such that single-particle
contributions are eliminated \cite{JaegerSV,JaegerHS}. $p\left( \left\vert
x\right\rangle _{1}|y\rangle _{2}\right) $ denotes the probabilities of
joint detections. As the visibilities explicitly depend on the form of the
transducers involved and the details of the measurement (e.g., the
measurement basis $\{\left\vert K\right\rangle $, $\left\vert L\right\rangle
\}$ is chosen as $\{\left\vert 0\right\rangle $, $\left\vert 1\right\rangle
\}$), we use the symbols $\mathcal{V}_{k}$, $\mathcal{V}_{12}$ here, to
indicate the experimental visibilities under a specific experimental
configuration, as opposed to the maximal visibilities $V_{k}$, $V_{12}$.

The \textquotedblleft corrected\textquotedblright\ two-particle joint
probabilities can be calculated as 
\begin{align}
\overline{p}\left( \left\vert x\right\rangle _{1}|y\rangle _{2}\right) & =%
\frac{1}{4}\{1+\left( -1\right) ^{x+y}[\left\vert M\right\vert \cos \left(
\phi _{1}+\phi _{2}-\xi _{1}\right)  \notag \\
& \qquad \qquad \qquad \quad +\left\vert N\right\vert \cos \left( \phi
_{1}-\phi _{2}-\xi _{2}\right) ]\},
\end{align}%
where%
\begin{equation}
\begin{array}{l}
M=\gamma _{1}\gamma _{4}^{\ast }-\left( \gamma _{1}\gamma _{3}^{\ast
}+\gamma _{2}\gamma _{4}^{\ast }\right) \left( \gamma _{1}\gamma _{2}^{\ast
}+\gamma _{3}\gamma _{4}^{\ast }\right) =\left\vert M\right\vert e^{i\xi
_{1}}, \\ 
N=\gamma _{2}\gamma _{3}^{\ast }-\left( \gamma _{1}\gamma _{3}^{\ast
}+\gamma _{2}\gamma _{4}^{\ast }\right) \left( \gamma _{1}^{\ast }\gamma
_{2}+\gamma _{3}^{\ast }\gamma _{4}\right) =\left\vert N\right\vert e^{i\xi
_{2}},%
\end{array}
\label{e.MN}
\end{equation}%
The maximal and minimal values of $\overline{p}\left( \left\vert
x\right\rangle _{1}|y\rangle _{2}\right) $ are thus 
\begin{equation}
\overline{p}_{\max ,\min }\left( \left\vert x\right\rangle _{1}|y\rangle
_{2}\right) =\frac{1}{4}\left[ 1\pm 2\left( \left\vert M\right\vert
+\left\vert N\right\vert \right) \right] .
\end{equation}%
These values are reached only when the phase shifters are set to $\left(
\phi _{1},\phi _{2}\right) =\left( n\pi +\frac{\xi _{1}+\xi _{2}}{2},m\pi +%
\frac{\xi _{1}-\xi _{2}}{2}\right) $, where the parameters $n,m$ can be $%
\left( n,m=0,\pm 1\right) $. Hence, on substituting for the maximal and
minimal values of these probabilities in Eq. (\ref{def.v12}), we find 
\begin{equation}
\mathcal{V}_{12}=2\left( \left\vert M\right\vert +\left\vert N\right\vert
\right) .  \label{e.v12}
\end{equation}%
With Eqs. (\ref{e.vis}), (\ref{e.MN}), and (\ref{e.v12}), the
complementarity relation (\ref{e.vv}) is obtained, valid for arbitrary pure
bipartite states.

\subsection{\protect\bigskip Experiments on two extreme cases}

For the experimental measurements, we used the nuclear spins of $^{13}$%
C-labeled chloroform as a representative 2-qubit quantum system. We
identified the spin of the $^{1}$H nuclei with particle 1 and the carbon
nuclei ($^{13}$C) with particle 2. The spin-spin coupling constant $J$
between $^{13}$C and $^{1}$H is 214.95 Hz. The relaxation times were
measured to be $T_{1}=16.5$ $sec$ and $T_{2}=6.9$ $sec$ for the proton, and $%
T_{1}=21.2$ $sec$ and $T_{2}=0.35$ $sec$ for the carbon nuclei. Experiments
were performed on an Infinity+ NMR spectrometer equipped with a Doty probe
at the frequencies 150.13MHz for $^{13}$C and at 599.77MHz for $^{1}$H,
using conventional liquid-state NMR techniques.

For most of the experiments that we discuss in the following, the system was
first prepared into a pseudo-pure state $\rho _{00}=\frac{1-\epsilon}{tr(%
\mathbf{1})}\mathbf{1}+\varepsilon |00\rangle \langle 00|$. Here, $\mathbf{1}
$ is the unity operator and $\varepsilon $ a small constant of the order of $%
10^{-5}$ determined by the thermal equilibrium. We used the spatial
averaging technique \cite{PPS} and applied the pulse sequence: 
\begin{equation}
\left[ \frac{\pi }{3}\right] _{\pi /2}^{1}-G_{z}-\left[ \frac{\pi }{4}\right]
_{\pi /2}^{1}-\frac{\pi /2}{\pi J}-\left[ \frac{\pi }{4}\right]
_{0}^{1}-G_{z} ,
\end{equation}
where $G_{z}$ is a field gradient pulse that destroys the transverse
magnetizations. The upper indices of the pulses indicate to which qubit the
rotation is applied.

Starting from this pseudo-pure state, we then prepared the two-particle
source states $\left\vert \Theta \right\rangle $. As an example, we consider
a product state $|\Phi \rangle $ 
\begin{equation}
\left\vert \Phi \right\rangle =\left[ \frac{1}{\sqrt{2}}\left( \left\vert
0\right\rangle _{1}+\left\vert 1\right\rangle _{1}\right) \right] \otimes %
\left[ \frac{1}{\sqrt{2}}\left( \left\vert 0\right\rangle _{2}+\left\vert
1\right\rangle _{2}\right) \right] ,
\end{equation}%
and a maximally entangled state $|\Psi \rangle $%
\begin{equation}
\left\vert \Psi \right\rangle =\frac{1}{\sqrt{2}}\left( \left\vert
0\right\rangle _{1}\left\vert 0\right\rangle _{2}+\left\vert 1\right\rangle
_{1}\left\vert 1\right\rangle _{2}\right) .
\end{equation}%
They can be prepared from $\rho_{00}$ by the following pulse sequences : 
\begin{equation}
\begin{array}{l}
\left\vert \Phi \right\rangle :\left[ \frac{\pi }{2}\right] _{\pi /2}^{1}%
\left[ \frac{\pi }{2}\right] _{\pi /2}^{2} \\ 
\left\vert \Psi \right\rangle :\left[ \frac{\pi }{2}\right] _{-\pi /2}^{1}%
\left[ \frac{\pi }{2}\right] _{-\pi }^{1}\left[ \frac{\pi }{2}\right] _{\pi
/2}^{1}\left[ \frac{\pi }{2}\right] _{-\pi }^{2}\left[ \frac{\pi }{2}\right]
_{\pi /2}^{2}-\frac{\pi /2}{\pi J}-\left[ \frac{\pi }{2}\right] _{\pi /2}^{2}%
\end{array}%
\end{equation}%
where $-\frac{\theta }{\pi J}-$ represents a free evolution for this time
under the scalar coupling.

The actual interferometer was realized by applying the transducers $\mathcal{%
U}\left( \phi _{1},\phi_{2}\right) $ of Eq. (\ref{U.op}) to the prepared
state $\left\vert \Theta \right\rangle $, which describes the effect of the
phase shifters and symmetric beam splitters. The transducer pulse sequence (%
\ref{pulse.Uk}) is simultaneously applied to both qubits.

\begin{figure}[tbh]
\begin{center}
\includegraphics[width = 0.8\columnwidth]{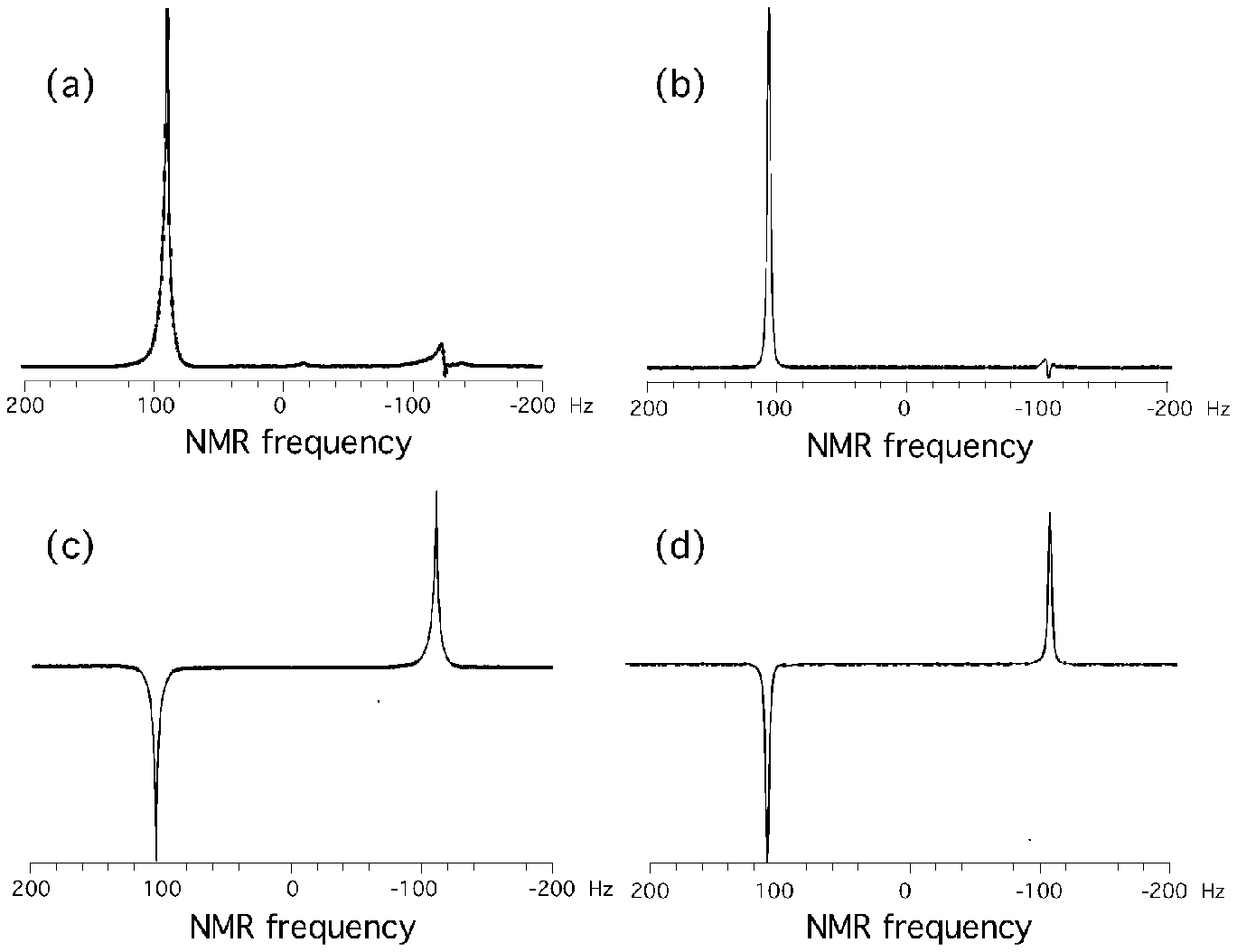}
\end{center}
\caption{Experimental spectra of proton and carbon at $\protect\phi _{k}=0$:
(a) and (b) for the product state $\left\vert \Phi \right\rangle $; (c) and
(d) for the entangled state $\left\vert \Psi \right\rangle $. (a),(c) are
the proton signals and (b),(d) are the carbon signals. }
\label{Spe1d}
\end{figure}

The probabilities that enter the complementarity relations can be expressed
in terms of populations of the four spin states. To determine these spin
states, we used a simplified quantum state tomography scheme to reconstruct
only the diagonal elements of the density matrix. This was realized by 
\begin{equation}
G_{z}-\left[ \frac{\pi }{2}\right] _{\pi /2}^{k}-FID_{k}
\end{equation}%
for $k=1,2$. $FID_{k}$ represents to recording the FID of qubit $k$ after a
field gradient pulse $G_{z}$ and a read-out pulse $\left[ \frac{\pi }{2}%
\right] _{\pi /2}^{k}$. Fig. \ref{Spe1d} shows the NMR signals after Fourier
transformation of the corresponding FIDs for the proton and carbon spins in $%
^{13}$CHCl$_{3}$ at $\phi _{k}=0$ when they are prepared in the product
state $|\Phi \rangle $ or the maximally entangled state $|\Psi \rangle $. 
The signals measure the populations: 
\begin{equation}
\begin{array}{c}
S_{NMR}\left( \text{Carbon}\right) \sim p(|0\rangle _{1}|x\rangle
_{2})-p(|1\rangle _{1}|x\rangle _{2}), \\ 
S_{NMR}\left( \text{Proton}\right) \sim p(|x\rangle _{1}|0\rangle
_{2})-p(|x\rangle _{1}|1\rangle _{2}),%
\end{array}%
\end{equation}%
where $x=0$ for the high-frequency resonance line and $1$ for the
low-frequency line.

\begin{figure}[tbh]
\begin{center}
\includegraphics[width = 0.8\columnwidth]{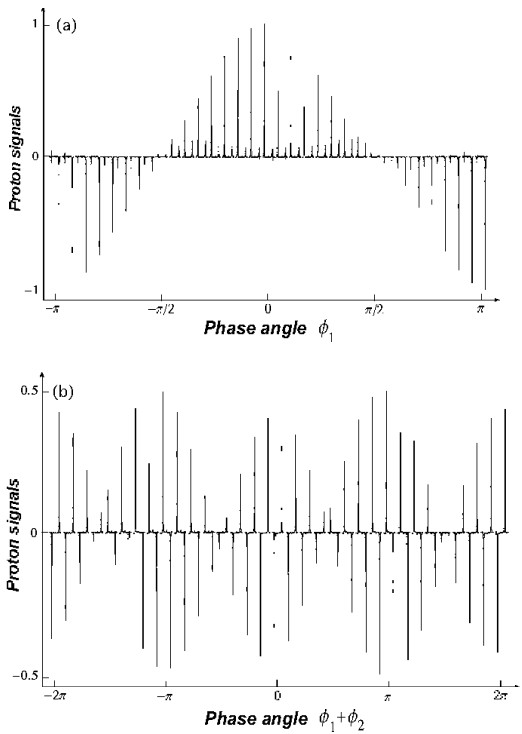}
\end{center}
\caption{Experimental spectra of proton with the phase $\protect\phi _{1}$:
(a) for the product state $\left\vert \Phi \right\rangle $ and (b) for the
entangled state $\left\vert \Psi \right\rangle $.}
\label{extr}
\end{figure}

To create an interferogram, we varied the phases $\phi _{k}$ from $0$ to $%
2\pi $, incrementing both simultaneously in steps of $\pi /16$. The
resulting interference pattern of the proton is shown in Fig. \ref{extr}. 
The carbon signals have a similar behavior as a function of $\phi _{2}$ for
the states $\left\vert \Phi \right\rangle $ and $\left\vert \Psi
\right\rangle $, as Fig. \ref{extr} shows. From these experimental data
points, we calculated the probabilities $p(|x\rangle_{k})$ and $\overline{p}%
\left( \left\vert x\right\rangle _{1}|y\rangle _{2}\right) $ and fitted
those to a cosine function: $y=A\ast cos(x-x_{0})+B$. From the fitted values
of the amplitude $A$ and the offset $B$, we extracted the experimental
visibilities as $\mathcal{V}_{1}=1.04\pm 0.02,$ $\mathcal{V}_{2}=0.99\pm
0.01 $, $\mathcal{V}_{12}=0.05\pm 0.01$ for the product state $\left\vert
\Phi \right\rangle $ and $\mathcal{V}_{1}=0.03\pm 0.01,$ $\mathcal{V}%
_{2}=0.14\pm 0.01$, $\mathcal{V}_{12}=0.86\pm 0.02$ for the entangled state $%
\left\vert \Psi \right\rangle $\ by the definitions of Eqs.(\ref{def.v}) and
(\ref{def.v12}). As theoretically expected, the product state $\left\vert
\Phi \right\rangle $ shows one-particle interference fringes, but almost no
two-particle interference fringes, while the situation is reversed for the
entangled state $\left\vert \Psi \right\rangle $. It can also be seen that
the discrepancies from the theory is larger for the entangled state $%
\left\vert \Psi \right\rangle $ than for the product state $\left\vert \Phi
\right\rangle $. This is easily understood by realizing that the state
preparation is more complicated for the entangled state.

\section{''Which-way'' information in bipartite systems}

\subsection{Predictability}

For the same system, we can calculate the predictabilities, i.e. the
probabilities for correctly predicting which path the particle will take,
from the expectation value of the $\sigma _{z}^{(k)}$ observable on the
state $\left\vert \Theta \right\rangle $, i.e., $\mathcal{P}_{k}=\left\vert
\left\langle \Theta \right\vert \sigma _{z}^{\left( k\right) }\left\vert
\Theta \right\rangle \right\vert $: 
\begin{eqnarray}
\mathcal{P}_{1} &=&\left\vert |\gamma _{1}|^{2}+|\gamma _{2}|^{2}-|\gamma
_{3}|^{2}-|\gamma _{4}|^{2}\right\vert  \notag \\
\mathcal{P}_{2} &=&\left\vert |\gamma _{1}|^{2}-|\gamma _{2}|^{2}+|\gamma
_{3}|^{2}-|\gamma _{4}|^{2}\right\vert ,  \label{e.pred}
\end{eqnarray}%
where $\sigma _{z}^{(k)}$ is the \textit{z} component of the Pauli operator
on qubit $k$. $\mathcal{P}_{k}$ is thus the magnitude of the difference
between the probabilities that particle $k$ takes path $|0\rangle _{k}$ or
the other path $|1\rangle _{k}$.

For the experimental measurement of the predictability $\mathcal{P}_{k}$, we
measure the observable $\sigma _{z}^{(k)}$ by partial quantum state
tomography: a field gradient pulse destroys coherences and a readout pulse $%
\left[ \frac{\pi }{2}\right] _{\pi /2}^{k}$ converts $\sigma _{z}^{(k)}$
into $\sigma _{x}^{(k)}$, which is recorded as the FID. Upon Fourier
transformation, the integral of both lines yields $\langle \sigma
_{z}^{(k)}\rangle $, and its magnitude corresponds to the predictability $%
\mathcal{P}_{k}$.

\begin{figure}[tbh]
\begin{center}
\includegraphics[width = 0.8\columnwidth]{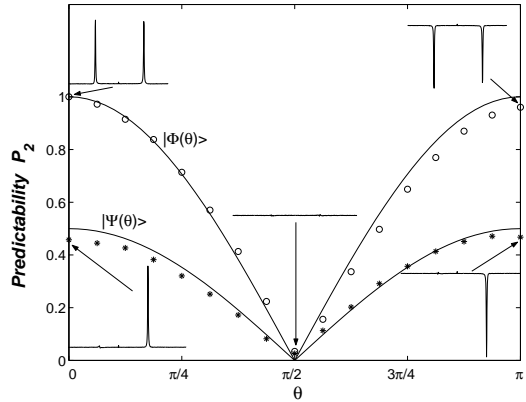}
\end{center}
\caption{Experimental meaurement of the predictability $\mathcal{P}_{2}$ for 
$\left\vert \Phi \left( \protect\theta \right) \right\rangle $ (denoted by $%
\bigcirc $) and $\left\vert \Psi \left( \protect\theta \right) \right\rangle 
$ (denoted by $\ast $). The solid lines are the theoretical expectations.
The insets are, respectively, the experimental spectra at $\protect\theta =0,%
\frac{\protect\pi }{2},\protect\pi $. }
\label{P}
\end{figure}

Figure \ref{P} shows the measurement of the predictability $\mathcal{P}_{2}$
on $^{13}$CHCl$_{3}$ for two specific examples: the product state $%
\left\vert \Phi \left( \theta \right) \right\rangle =\left[ \frac{1}{\sqrt{2}%
}\left( \left\vert 0\right\rangle _{1}+\left\vert 1\right\rangle _{1}\right) %
\right] \otimes \left[ \cos \frac{\theta }{2}\left\vert 0\right\rangle
_{2}+\sin \frac{\theta }{2}\left\vert 1\right\rangle _{2}\right] $ and the
entangled state $\left\vert \Psi \left( \theta \right) \right\rangle =\frac{1%
}{\sqrt{2}}\left( \left\vert 0\right\rangle _{1}\otimes \left[ \frac{1}{%
\sqrt{2}}(\left\vert 0\right\rangle _{2}+\left\vert 1\right\rangle _{2})%
\right] +\left\vert 1\right\rangle _{2}\otimes \left[ \cos \frac{\theta }{2}%
\left\vert 0\right\rangle _{2}+\sin \frac{\theta }{2}\left\vert
1\right\rangle _{2}\right] \right) $. 

\subsection{Distinguishability}

In a bipartite system, the which-way information for particle $k$ can be
optimized by first performing a projective measurement on particle $j\left(
j\neq k\right) $. For this measurement, we first have to choose the optimal
ancilla observable $W_{j}^{(opt)}$. According to Englert's quantitative
analysis of the distinguishability\cite{Englert}, we start by writing the
quantum state $|\Theta \rangle $ as the sum of two components corresponding
to two paths of qubit $k$: 
\begin{equation}
\left\vert \Theta \right\rangle =a_{k+}|0\rangle _{k}|m_{+}\rangle
_{j}+a_{k-}|1\rangle _{k}|m_{-}\rangle _{j}.
\end{equation}%
Each component is coupled to a different state of qubit $j$: 
\begin{equation}
\begin{array}{c}
|m_{+}\rangle _{j}=\frac{\gamma _{1}}{a_{k+}}|0\rangle _{j}+\frac{\gamma
_{1+k}}{a_{k+}}|1\rangle _{j} \\ 
|m_{-}\rangle _{j}=\frac{\gamma _{4-k}}{a_{k-}}|0\rangle _{j}+\frac{\gamma
_{4}}{a_{k-}}|1\rangle _{j}%
\end{array}%
.  \label{e.m}
\end{equation}%
The coefficients $a_{k\pm }$ are 
\begin{equation}
\begin{array}{c}
a_{k+}=\sqrt{|\gamma _{1}|^{2}+|\gamma _{1+k}|^{2}} \\ 
a_{k-}=\sqrt{|\gamma _{4-k}|^{2}+|\gamma _{4}|^{2}}%
\end{array}%
.  \label{e.ak}
\end{equation}%
A suitable measurement is performed on qubit $j$ to make qubit $k$ acquire
the maximal \textquotedblleft which-way" information. To determine the most
useful ancilla observable, we write it as $W_{j}=\vec{b}\cdot \vec{\sigma}%
^{(j)}$. The probability that the ancilla observable finds eigenvalue $%
\lambda $ differs for the two component states: 
\begin{equation}
\begin{array}{c}
p_{+}(\lambda )=a_{k+}^{2}\langle \psi _{\lambda }|(|m_{+}\rangle
_{j}\langle m_{+}|_{j})|\psi _{\lambda }\rangle \\ 
p_{-}(\lambda )=a_{k-}^{2}\langle \psi _{\lambda }|(|m_{-}\rangle
_{j}\langle m_{-}|_{j})|\psi _{\lambda }\rangle%
\end{array}%
,
\end{equation}%
where $\psi _{\lambda }$ is the corresponding eigenvector.

The distinguishability $D_{k}$ for qubit $k$ is obtained by maximizing the
difference of the measurement probabilities for the two components, 
\begin{equation}
D_{k}=\max \left\{ \sum_{\lambda }|p_{+}(\lambda )-p_{-}(\lambda )|\right\} .
\label{e.Dkm}
\end{equation}%
Using the notation 
\begin{equation}
|m_{\pm }\rangle \langle m_{\pm }|=\vec{m_{\pm }}\cdot \vec{\sigma},
\label{e.mv}
\end{equation}%
where $\vec{m_{\pm }}$ are vectors on the Bloch sphere, we write 
\begin{equation}
D_{k}=\max \left\{ \left\vert \vec{b}\cdot \left[ a_{k+}^{2}\vec{m}%
_{+}-a_{k-}^{2}\vec{m}_{-}\right] \right\vert \right\} .  \label{e.Dk}
\end{equation}%
Clearly the maximum is reached if \emph{the two vectors }$\vec{b}$ \emph{and}
$\left[ a_{k+}^{2}\vec{m}_{+}-a_{k-}^{2}\vec{m}_{-}\right] $\emph{\ are} 
\emph{parallel}. Since $\vec{b}$ has unit length, the distinguishability
becomes 
\begin{eqnarray}
D_{k} &=&\left\Vert a_{k+}^{2}\vec{m}_{+}-a_{k-}^{2}\vec{m}_{-}\right\Vert  
\notag \\
&=&\sqrt{1-2a_{k+}^{2}a_{k-}^{2}\left( 1+\vec{m}_{+}\cdot \vec{m}_{-}\right) 
}.  \label{e.D}
\end{eqnarray}%
Combining Eqs. (\ref{e.vis}) and (\ref{e.D}), we obtain the complementarity
relation (\ref{e.dvp}), i.e., $D_{k}^{2}+V_{k}^{2}=1$ for $k=1,2$.

For the experimental measurement, we first have to perform a "measurement"
on the ancilla qubit $j$, using the optimal observable $W_{j}=\vec{b}\cdot 
\vec{\sigma _{j}}$. This is done by applying a unitary transformation $R$ to
rotate the eigenbasis $\{|\psi _{\lambda }\rangle _{j}\}$ of the observable $%
W_{j}^{(opt)}$ into the computational basis $\{|0\rangle _{j},|1\rangle
_{j}\}$ \cite{Brassard}. The subsequent field gradient pulse destroys
coherence of qubit $j$ \cite{Teklemariam}, as well as qubit $k$ and joint
coherences (=zero and double quantum coherences). After this ancilla
measurement, the distinguishability $D_{k}$ can be measured by a readout
pulse, detection of the FID, Fourier transformation, and taking the sum of
the magnitudes of both resonance lines.

\begin{figure}[tbh]
\begin{center}
\includegraphics[width = 0.8\columnwidth]{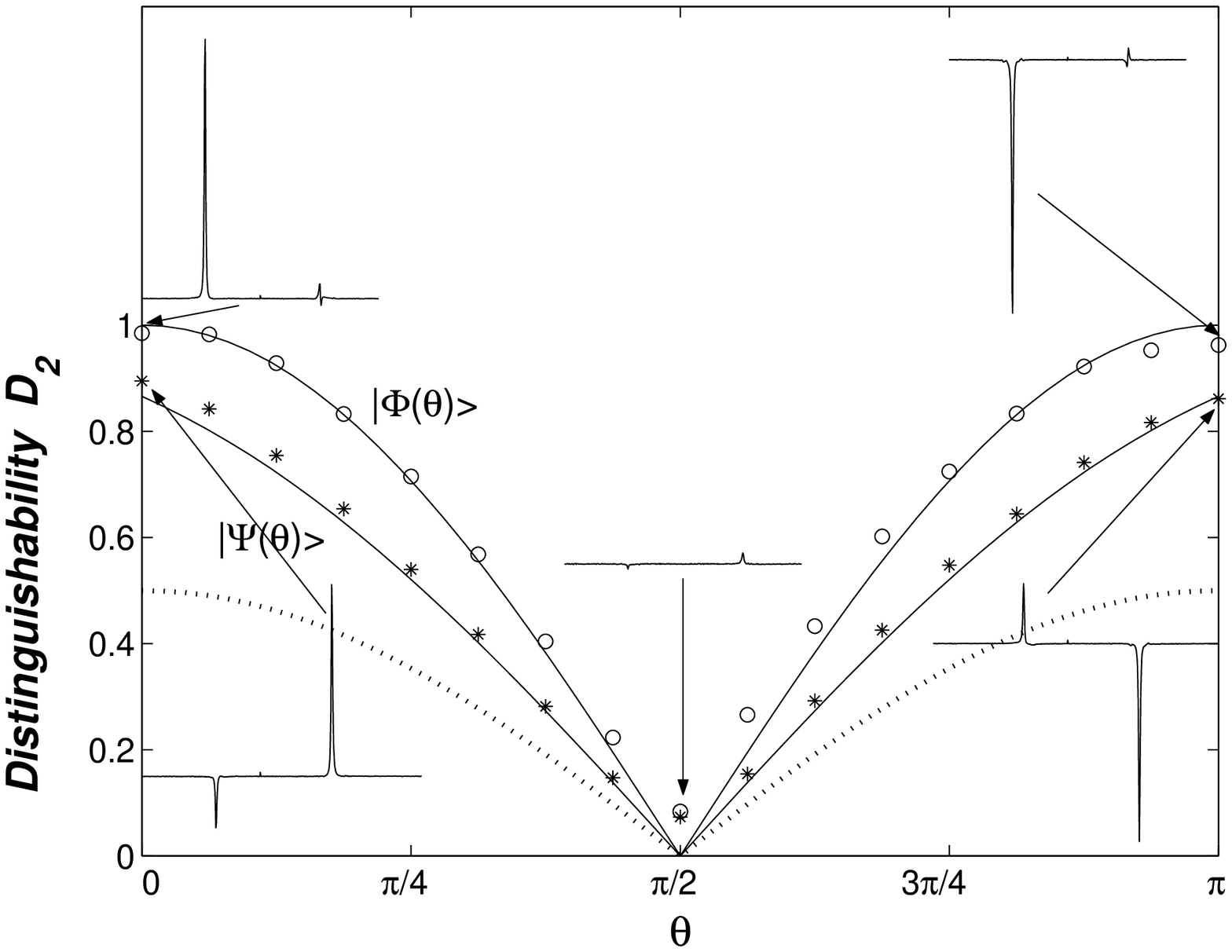}
\end{center}
\caption{Experimental meaurement of the distinguishability $D_{2}$ for $%
\left\vert \Phi \left( \protect\theta \right) \right\rangle $ (denoted by $%
\bigcirc $) and $\left\vert \Psi \left( \protect\theta \right) \right\rangle 
$ (denoted by $\ast $). The solid lines are the theoretical expectations and
the dotted line is the theoretical expectation of the predictability $%
\mathcal{P}_{2}$ for $\left\vert \Psi \left( \protect\theta \right)
\right\rangle $. The insets are, respectively, the experimental spectra at $%
\protect\theta =0,\frac{\protect\pi }{2},\protect\pi $. }
\label{D}
\end{figure}

Figure \ref{D} shows the observed distinguishability $D_{2}$ for the states $%
\left\vert \Phi \left( \theta \right) \right\rangle $ and $\left\vert \Psi
\left( \theta \right) \right\rangle $. From Eq. (\ref{e.Dk}), we find the
optimal observable $W_{1}^{(opt)}$ is $\sigma _{z}^{(1)}$ for $\left\vert
\Phi \left( \theta \right) \right\rangle $, and $\sin (\kappa /2)\sigma
_{x}^{(1)}+\cos (\kappa /2)\sigma _{z}^{(1)}$ with $\kappa =\arctan (-\sec (%
\frac{\pi }{4}-\frac{\theta }{2}))$ for $\left\vert \Psi \left( \theta
\right) \right\rangle $. Therefore, the transformation $R$ was realized by
the NMR pulses$\ \left[ \frac{\pi }{2}\right] _{\frac{3\pi }{2}}^{1}$ and $%
\left[ \kappa \right] _{\frac{3\pi }{2}}^{1}$. As there is no entanglement
in the state $\left\vert \Phi \left( \theta \right) \right\rangle $, $D_{2}=%
\mathcal{P}_{2}$, while for the entangled state $\left\vert \Psi \left(
\theta \right) \right\rangle $ we find $D_{2}>\mathcal{P}_{2}$. The
experimental data also satisfy the relation $D_{2}^{2}=\mathcal{P}%
_{2}^{2}+C^{2}$ of Eq. (\ref{D_pc}). 

\section{Complementarity relations for bipartite systems}

With the same experimental scheme we now explore the complementarity
relations for bipartite quantum systems. Between the single particle
visibility $\mathcal{V}_{k}$ (see Eq. (\ref{e.vis})), the two-particle
visibility $\mathcal{V}_{12}$ (Eq. (\ref{e.v12})), and the predictability $%
\mathcal{P}_{k}$ (Eq. (\ref{e.pred})), we can verify that the relation 
\begin{equation}
\mathcal{V}_{12}^{2}+\mathcal{V}_{k}^{2}+\mathcal{P}_{k}^{2}\leq 1\quad
\left( k=1,2\right) ,  \label{e.vvp}
\end{equation}%
holds in a pure bipartite system for any experimental setting and
measurement basis.

If the initial state $\left\vert \Theta \right\rangle $ has only real
coefficients $\gamma _{i}$, the inequality becomes an equality. In this
case, the two-particle visibility $\mathcal{V}_{12}$\ becomes equal to the
concurrence $C$, $\mathcal{V}_{12}\equiv C=2\left\vert \gamma _{1}\gamma
_{4}-\gamma _{2}\gamma _{3}\right\vert $. However, when the coefficients $%
\gamma _{i}$ are arbitrary complex numbers, the two-particle visibility $%
\mathcal{V}_{12}$ can be smaller than the concurrence, $\mathcal{V}_{12}\leq
C$. As a specific example consider $\left\vert \Theta \right\rangle
=-0.3\left\vert 00\right\rangle -0.2e^{-i\frac{3\pi }{5}}\left\vert
01\right\rangle +0.8e^{-i\frac{\pi }{25}}\left\vert 10\right\rangle
+0.4796e^{-i\frac{5\pi }{12}}\left\vert 11\right\rangle $. Using symmetric
beam splitters and the measurement basis $\{\left\vert 0\right\rangle
,|1\rangle \}{}$), we find $\mathcal{V}_{12}=0.1627$ and $C=0.2110$, i.e. $%
\mathcal{V}_{12}<C$.

By the Schmidt decomposition\cite{Nielsbook}, any pure state $\left\vert
\Theta \right\rangle $ can be transformed into one with real coefficients by
local unitary operations. Therefore, one can design a different experiment
using beam splitters that implement the transformation $e^{i\frac{\alpha _{k}%
}{2}\left( \sigma _{x}^{(k)}\cos\xi_k +\sigma _{y}^{(k)}\sin \xi_k \right) }$
instead of the symmetric one $e^{i\frac{\pi }{4}\sigma _{y}^{(k)}}$. In this
case, the single-particle transducers implement the operation 
\begin{equation}
U_{k}\left( \theta _{k},\xi _{k},\phi _{k}\right) = e^{i\frac{\alpha _{k}}{2}%
\left( \sigma _{x}^{(k)}\cos \xi_k +\sigma _{y}^{(k)}\sin \xi_k \right)
}e^{-i\frac{\phi _{k}}{2}\sigma _{z}^{(k)}}
\end{equation}%
instead of $U_{k}\left( \phi _{k}\right) $ in Eq. (\ref{Uk.op}). Note that
the single-particle character $S_{k}$ (Eq. (\ref{e.Sk})) is invariant under
local unitary transformations though its constituents $\mathcal{V}_{k}$ and $%
\mathcal{P}_{k}$ are not. By defining the maximal visibility $%
V_{12}=\max_{\{\alpha _{k},\xi _{k}\}}\left\{ \mathcal{V}_{12}\left( \alpha
_{k},\xi _{k}\right) \right\} $, we obtain $V_{12}\equiv C$ and 
\begin{equation}
V_{12}^{2}+S_{k}^{2}=1,\quad \left( k=1,2\right) .  \label{V12S}
\end{equation}%
This shows that the complementarity relation (\ref{e.Vvp}) in the equality
form is fulfilled for any pure bipartite system. An alternative way is to
keep the symmetric beam splitters and change the measurement basis. One can
always choose an optimal basis which consists of the eigenvectors of an
observable $W=W_{1}\otimes W_{2}$ that maximizes the visibility $\mathcal{V}%
_{12}$, i.e., $V_{12}=\max_{\{W\}}\left\{ \mathcal{V}_{12}\left( W\right)
\right\} \equiv C$. Being invariant under local unitary transformations,
this maximal two-particle visibility $V_{12}$ (= concurrence $C$) is a good
measure of the bipartite property encoded in the pure state.

In a pure bipartite system, the complementarity relation (\ref{V12S}),
together with the identity $V_{12}\equiv C$ and the definition (\ref{e.Sk})
of the single particle character $S_{k}$ offers a method for quantifying
entanglement in terms of the directly measurable quantities, in this case
visibilities, predictability and distinguishability. In this section, we
experimentally explore these complementarity relations for the states 
\begin{eqnarray}
\left\vert \psi \left( \theta _{1},\theta _{2}\right) \right\rangle &=&\frac{%
1}{\sqrt{2}}[\left\vert 0\right\rangle _{1}\otimes \left( \cos \frac{\theta
_{1}}{2}\left\vert 0\right\rangle _{2}+\sin \frac{\theta _{1}}{2}\left\vert
1\right\rangle _{2}\right) \newline
\notag \\
&&+\left\vert 1\right\rangle _{1}\otimes \left( \cos \frac{\theta _{2}}{2}%
\left\vert 0\right\rangle _{2}+\sin \frac{\theta _{2}}{2}\left\vert
1\right\rangle _{2}\right) ]  \label{spec.state}
\end{eqnarray}%
by preparing the state in the nuclear spins of molecules, and measuring the
visibilities, predictability and distinguishability by NMR according to the
procedure outlined above.

\begin{table*}[tbp]
\begin{tabular}{c|c|c|c}
\hline\hline
Particle $k$ & $C\equiv V_{12}$ & $S_{k},%
\begin{array}{c}
V_{k} \\ 
P_{k}%
\end{array}%
$ & $D_{k}$ \\ \hline
\multicolumn{1}{l|}{%
\begin{tabular}{l}
1$%
\begin{array}{c}
~ \\ 
~%
\end{array}%
$ \\ \hline
2$%
\begin{array}{c}
~ \\ 
~%
\end{array}%
$%
\end{tabular}%
} & \multicolumn{1}{|l|}{$\left\vert \sin \left( \frac{\theta _{1}-\theta
_{2}}{2}\right) \right\vert $} & \multicolumn{1}{|l|}{%
\begin{tabular}{l}
$\left\vert \cos \left( \frac{\theta _{1}-\theta _{2}}{2}\right) \right\vert
,%
\begin{array}{c}
\left\vert \cos \left( \frac{\theta _{1}-\theta _{2}}{2}\right) \right\vert
\\ 
0%
\end{array}%
$ \\ \hline
$\left\vert \cos \left( \frac{\theta _{1}-\theta _{2}}{2}\right) \right\vert
,%
\begin{array}{c}
\left\vert \sin \left( \frac{\theta _{1}+\theta _{2}}{2}\right) \cos \left( 
\frac{\theta _{1}-\theta _{2}}{2}\right) \right\vert \\ 
\left\vert \cos \left( \frac{\theta _{1}+\theta _{2}}{2}\right) \cos \left( 
\frac{\theta _{1}-\theta _{2}}{2}\right) \right\vert%
\end{array}%
$%
\end{tabular}%
} & \multicolumn{1}{|l}{%
\begin{tabular}{l}
$\left\vert \cos \left( \frac{\theta _{1}-\theta _{2}}{2}\right) \right\vert 
\begin{array}{c}
~ \\ 
~%
\end{array}%
$ \\ \hline
$\sqrt{1-\sin ^{2}(\frac{\theta _{1}+\theta _{2}}{2})\cos ^{2}\left( \frac{%
\theta _{1}-\theta _{2}}{2}\right) }%
\begin{array}{c}
~ \\ 
~%
\end{array}%
$%
\end{tabular}%
} \\ \hline\hline
\end{tabular}%
\caption{The various quantities involved in the complementarity relation for
the family of states $\left\vert \protect\psi \left( \protect\theta _{1},%
\protect\theta _{2}\right) \right\rangle $ in Eq. (\protect\ref{spec.state}%
). }
\label{tab.inter}
\end{table*}

Table \ref{tab.inter} lists the theoretical expectations for the various
quantities involved in the complementarity relation for this state. The
single particle character $S_{k}$ and the concurrence $C$ ($\equiv $ the
maximal two-particle visibility $V_{12}$) satisfy the duality relation of
Eq. (\ref{V12S}). For the state $\left\vert \psi \left( \theta _{1},\theta
_{2}\right) \right\rangle $ (Eq. (\ref{spec.state})), the maximal
two-particle visibility $V_{12}$ is obtained from the experimental
visibility $\mathcal{V}_{12}$ by setting the measurement basis $\left\{
|x_{1}y_{2}\rangle \right\} $ to the computational basis $\left\{ |x,y=0%
\text{ }or\text{ }1\rangle \right\} $. The predictabilities for the two
particles are qualitatively different, $P_{1}\neq P_{2}$, which results in $%
V_{1}^{2}+V_{12}^{2}=1$ whereas $V_{2}^{2}+V_{12}^{2}\leq 1$. The special
case with $\theta _{1}+\theta _{2}=\pi $ was discussed in detail in Ref.\cite%
{JaegerHS}; in that case, both predictabilities vanish, $P_{1}=P_{2}=0$.
However, $V_{12}^{2}+V_{k}^{2}+P_{k}^{2}=1$ and $D_{k}^{2}+V_{k}^{2}=1$ are
still satisfied for $k=1$ or $2$.

To verify these relations, we used an experimental procedure similar to that
discussed in Section IIIB. To prepare the state $\left\vert \psi \left(
\theta _{1},\theta _{2}\right) \right\rangle $ from the pseudo-pure state $%
\rho _{00}$, we used the following NMR pulse sequence: 
\begin{equation}
\left\vert \psi \left( \theta _{1},\theta _{2}\right) \right\rangle :\left[ 
\frac{\pi }{2}\right] _{\pi /2}^{1}\left[ \frac{\pi }{2}\right] _{\pi }^{2}-%
\frac{\theta _{1}-\theta _{2}}{2\pi J}-\left[ \frac{\pi }{2}\right] _{-\pi
}^{2}\left[ \frac{\theta _{1}+\theta _{2}}{2}\right] _{\pi /2}^{2}.
\end{equation}%
When $(\theta _{1}-\theta _{2})/2\pi J$ is negative, we generate the
required evolution by inserting two $\pi $ pulses on one of the two qubits
before and after the evolution period of $\left\vert \theta _{1}-\theta
_{2}\right\vert /2\pi J$.

We measured the visibilities for the state $\left\vert \psi \left( \theta
_{1},\theta _{2}\right) \right\rangle $ by first scanning $\phi _{1}$ while
fixing $\phi _{2}$ to $\pi /2$ , then repeated the experiment with fixed $%
\phi _{1}$ and variable $\phi _{2}$. This provides the maximal probabilities 
$\overline{p}_{\max ,\min }\left( \left\vert x\right\rangle _{1}|y\rangle
_{2}\right) $ of the \textquotedblleft corrected\textquotedblright\ joint
probabilities, which occur at $(\phi _{1},\phi _{2})=\left( n\pi +\frac{\pi 
}{2},m\pi +\frac{\pi }{2}\right) $ for $\left\vert \psi \left( \theta
_{1},\theta _{2}\right) \right\rangle $.

\begin{figure}[tbh]
\begin{center}
\includegraphics[width = 0.9\columnwidth]{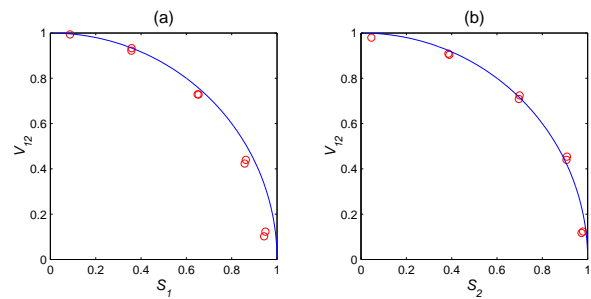}
\end{center}
\caption{(Color online) Experimental verification of the complementarity
relation $V_{12}^{2}+S_{k}^{2}=1$ in a pure two-qubit system: (a) for qubit
1 and (b) for qubit 2. Solid curves represent the theoretical
complementarity relation of single-particle character $S_k$ versus
two-particle visibility $V_{12}$. Experimental results are indicated by
circles. }
\label{S.V12}
\end{figure}

As a specific example, we present the experimental results for $\left\vert
\psi \left( \theta _{1},\theta _{2}\right) \right\rangle $ with $\theta
_{1}+\theta _{2}=\frac{\pi }{2}$. The resulting interference fringes were
closely similar to those shown in Fig. \ref{extr}. Using the procedure
described in Section IIIB, we extracted the relevant visibilities $V_{k}$
and $V_{12}$ from the experimental data. The visibilities and the
predictability were measured as a function of $\theta _{1}$ varying from $%
-\pi /4$ to $3\pi /4$ in steps of $\pi /8$. The single-particle character $%
S_{k}=\sqrt{V_{k}^{2}+P_{k}^{2}}$ and the two-particle visibility $V_{12}$
from these experiments are displayed in Fig. \ref{S.V12}, together with
plots of the theoretical complementarity relations (solid curves) indicating 
$V_{12}^{2}+S_{k}^{2}=1$ for the pure two-qubit states. A fit of these data
to the equation $x^{2}+y^{2}=r^{2}$ resulted in an amplitude $r=0.98\pm 0.01$
for the data of Fig. \ref{S.V12} (a) and $0.97\pm 0.01$ for the data of Fig. %
\ref{S.V12}(b).

\begin{figure}[tbh]
\begin{center}
\includegraphics[width = 0.9\columnwidth]{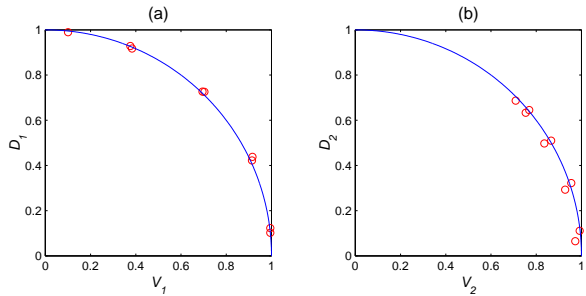}
\end{center}
\caption{(Color online) Experimental verification of the complementarity
relation of $D_{k}^{2}+V_{k}^{2}=1$ in a pure two-qubit system. (a) for
qubit 1 and (b) for qubit 2. Solid curves represent the ideal
complementarity relationship, while experimental results are indicated by
circles. }
\label{D.V}
\end{figure}

For the quantitative measurement of the distinguishability $D_{k}$, the
optimal observable $W_{j}^{(opt)}$ for $\left\vert \psi \left( \theta
_{1},\theta _{2}\right) \right\rangle $ is a spin operator parallel to $\vec{%
b}=(\sin (\kappa /2),0,\cos (\kappa /2))$ with $\kappa =\frac{\theta
_{1}+\theta _{2}}{2}-\frac{\pi }{2}$ for $D_{1}$, in agreement with Ref. 
\cite{Peng}, and $\kappa =\arctan (-\cot (\frac{\theta _{1}+\theta _{2}}{2}%
)/\sin (\frac{\theta _{1}-\theta _{2}}{2}))$ for $D_{2}$, according to the
analysis of section IV B. The transformation $R_{j}$ was realized by a $%
\left[ \theta \right] _{\frac{3\pi }{2}}^{j}$ pulse. Figure \ref{D.V}
compares the measured values of the single particle visibilities $V_{k}$ and
the distinguishabilities $D_{k}$ to the theoretical complementarity
relations (solid curves) $D_{k}^{2}+V_{k}^{2}=1$. The fitted values of the
amplitude $r$ are $0.99\pm 0.01$ for the data in Fig. \ref{D.V}(a) and $%
0.98\pm 0.01$ for Fig. \ref{D.V}(b).

\begin{figure}[tbh]
\begin{center}
\includegraphics[width = 0.6\columnwidth]{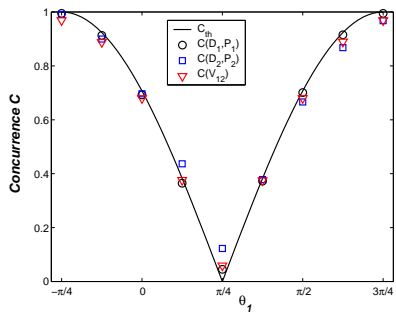}
\end{center}
\caption{(Color online) Measured concurrence from the experimental values of 
$V_{12}$ (denoted by $\bigtriangledown $) and $\protect\sqrt{%
D_{k}^{2}-P_{k}^{2}}$ (denoted, respectively, by $\bigcirc $ and $\boxdot $
for $k=1$ and $2$) verus $\protect\theta _{1}$. The solid curve represent
the theoretical concurrence $C=\left\vert \sin (\protect\theta _{1}-\frac{%
\protect\pi }{4})\right\vert $.}
\label{C_exp}
\end{figure}

In Fig. \ref{C_exp}, we compare two independent ways for measuring the
concurrence $C$, either through the two-particle visibility $V_{12}$, or
through the single-particle quantities, as $\sqrt{D_{k}^{2}-P_{k}^{2}}$.
Both data sets are plotted against $\theta _{1}$, together with the
theoretical concurrence $C$. The figure shows clearly that the two procedure
give the same results, within experimental errors. Apparently, both methods
allow one to experimentally determine the entanglement of pure two-qubit
states. At the same time, the data verify the complementarity relation (\ref%
{e.cvp}).

In these experiments, the maximal absolute errors for the quantities $V_{k}$%
, $V_{12}$ and $P_{k} $ were about $0.1$. The error is primarily due to the
inhomogeneity of the radio frequency field and the static magnetic field,
imperfect calibration of radio frequency pulses, and signal decay during the
experiments. A maximal experimental error about $6\%$ results for the
verification of the complementarity relations. If we take into account these
imperfections, the measured data in our NMR experiments agree well with the
theory.

\section{Multi-qubit systems}

To generalize the complementarity relation (\ref{e.cvp}) to multi-qubit
systems, we consider a pure state $\left\vert \psi \right\rangle $ with 
\textit{n} qubits $i,j,k,...,m$. According to the generalized concurrence
for pairs of quantum systems of arbitrary dimension by Rungta \textit{et al}%
. \cite{RBCHM01,RC2003}, we calculate the bipartite concurrence $%
C_{k(ij...m)}$ between qubit $k$ and the system with the remaining $n-1$
qubits $(ij...m)$ in terms of the marginal density operator $\rho _{k}$ 
\begin{equation}
C_{k(ij...m)}=\sqrt{2\left[ 1-Tr\left( \rho _{k}^{2}\right) \right] }.
\end{equation}%
In terms of the single particle character $S_{k}$ (Eq. (\ref{e.Sk})), and
using $Tr\left( \rho _{k}^{2}\right) =\frac{1}{2}\left( 1+S_{k}^{2}\right) $%
, we obtain the complementarity relation 
\begin{equation}
C_{k(ij...m)}^{2}+S_{k}^{2}=1  \label{e.nqubit}
\end{equation}%
This is a first generalization of the tradeoff between individual particle
properties, quantified by $S_{k}$, and the bipartite entanglement $%
C_{k(ij...m)}$ to many particle systems. It implies also the relation $%
\sum\limits_{i=1}^{n}\left[ C_{k(ij...m)}^{2}+S_{k}^{2}\right] =n$ derived
by Tessier\cite{Tessier}.

To characterize the pairwise entanglements of qubit $k$ with the other
qubits, we sum over the squares of the concurrences of all two-partite
subsystems involving qubit $k$, 
\begin{equation}
\tau _{2}^{(k)}=\sum\limits_{j\neq k}C_{kj}^{2}.  \label{e.tau2}
\end{equation}%
Here, the concurrence $C_{kj}$ is defined in terms of the marginal density
operator $\rho _{kj}$ for the $kj$ subsystem, using the definition of $%
C(\rho _{kj})=max\{\lambda _{1}-\lambda _{2}-\lambda _{3}-\lambda _{4},0\}$,
where $\lambda _{i}(i=1,2,3,4)$ are the square roots of the eigenvalues of $%
\rho _{kj}(\sigma _{y}^{(k)}\sigma _{y}^{(j)})\rho _{kj}^{\ast }(\sigma
_{y}^{(k)}\sigma _{y}^{(j)})$ in decreasing order\cite%
{Wootters1998,Wootters2001}.

We now specialize to pure three-qubit systems. Here, it is possible to
specify three-partite entanglement by the 3-tangle $\tau _{3}$\cite{CKW2000}
as 
\begin{equation}
\tau _{3}=C_{k(ij)}^{2}-C_{ki}^{2}-C_{kj}^{2}.  \label{def.tau3}
\end{equation}%
Combining Eqs (\ref{e.nqubit}), (\ref{e.tau2}), and (\ref{def.tau3}), we
find a complementarity between single-particle properties $S_{k}$, pair-wise
entanglement $\tau _{2}^{(k)}$, and three-partite entanglement $\tau _{3}$,
which is valid for each individual qubit: 
\begin{equation}
\tau _{3}+\tau _{2}^{(k)}+S_{k}^{2}=1,  \label{e.123}
\end{equation}

For specific examples, we have listed in Table \ref{t.states} different
three-qubit states and calculated the 1-, 2-, and 3-qubit quantifiers
appearing in Eq. (\ref{e.123}). As can be verified from the table, these
states satisfy Eq. (\ref{e.123}) in different ways. The product states of
the first entry only have single particle character. As discussed by D\"{u}r
et al. \cite{DVC2000}, the states listed in the second entry represent
bipartite entanglement between the second and third qubit, while the first
qubit is in a product state with them. The GHZ states are pure
three-particle entangled states, while the W states exhibit no \emph{genuine}
three-particle entanglement, but two- and one particle properties.

\begin{table*}[tbp]
\begin{tabular}{l|c|c|c}
\hline\hline
Class & $\tau _{3}$ & $\tau _{2}^{(k)}$ & $S_{k}^{2}=V_{k}^{2}+P_{k}^{2}$ \\ 
\hline
$%
\begin{array}{l}
\text{Product states}%
\end{array}%
$ & 0 & 0 & 1 \\ \hline
$%
\begin{array}{l}
\text{Bipartite entanglement} \\ 
\left\vert \psi _{r-st}\right\rangle =\left\vert 0\right\rangle _{r}\left(
a_{1}\left\vert 00\right\rangle +a_{2}\left\vert 11\right\rangle \right)
_{st}%
\end{array}%
$ & 0 & $%
\begin{array}{l}
0,(k=r) \\ 
4\left\vert a_{1}a_{2}\right\vert ^{2},\left( k=s,t\right)%
\end{array}%
$ & $%
\begin{array}{l}
1,\left( k=r\right) \\ 
(\left\vert a_{1}\right\vert ^{2}-\left\vert a_{2}\right\vert
^{2})^{2},\left( k=s,t\right)%
\end{array}%
$ \\ \hline
$%
\begin{array}{l}
\text{W states} \\ 
\left\vert W\right\rangle =a_{1}\left\vert 001\right\rangle +a_{2}\left\vert
010\right\rangle +a_{3}\left\vert 100\right\rangle%
\end{array}%
$ & 0 & $\sum\limits_{j\neq k}4\left\vert a_{k}a_{j}\right\vert ^{2}$ & $%
(\left\vert a_{k}\right\vert ^{2}-\sum\limits_{j\neq k}\left\vert
a_{j}\right\vert ^{2})^{2}$ \\ \hline
$%
\begin{array}{l}
\text{GHZ states} \\ 
\left\vert GHZ\right\rangle =a_{1}\left\vert 000\right\rangle
+a_{2}\left\vert 111\right\rangle%
\end{array}%
$ & $4\left\vert a_{1}a_{2}\right\vert ^{2}$ & 0 & $(\left\vert
a_{1}\right\vert ^{2}-\left\vert a_{2}\right\vert ^{2})^{2}$ \\ \hline\hline
\end{tabular}%
\caption{Some examples for the complementarity relation $\protect\tau _{3}+%
\protect\tau _{2}^{(k)}+S_{k}^{2}=1$ in a pure 3-qubit system.}
\label{t.states}
\end{table*}

Since there is no generalization of the 3-tangle to larger systems, we can
only speculate here if it is possible to extend the relation (\ref{e.123})
to more than three qubits. On a heuristic basis, we consider two types of
pure \textit{n}-qubit systems. One is a generalization of the GHZ states to 
\textit{n} qubits: $|GHZ_{n}\rangle =a_{1}|0\rangle ^{\otimes
n}+a_{2}|1\rangle ^{\otimes n}$. This is a state with pure \textit{n}-way
entanglement, i.e., 
\begin{eqnarray}
\tau _{n} &=&4|a_{1}a_{2}|^{2},\hspace{1cm}\tau _{m}^{(k)}=0\text{ for }1<m<n
\notag \\
S_{k}^{2} &=&\left( |a_{1}|^{2}-|a_{2}|^{2}\right) ^{2},
\end{eqnarray}%
where $\tau _{m}^{(k)}$ denotes the \textit{pure} \textit{m}-tangle
regarding qubit $k$. Here, the \textit{m}-tangle denotes \textit{m}-way or 
\textit{m}-party entanglement that critically involves all \textit{m}
parties, which is different from the I-tangle in Ref. \cite{RBCHM01,RC2003}
and a recently introduced measure of multi-partite entanglement defined by $%
C_{(n)}^{2}=Tr\left( \rho \tilde{\rho}\right) $ with a spin-flip operation $%
\tilde{\rho}\equiv \sigma _{y}^{\otimes n}\rho \sigma _{y}^{\otimes n}$ by
A. Wong and N. Christensen \cite{WC2001}. Currently, there is no general way
to measure this form of entanglement beyond three qubits.

The W states of Table \ref{t.states} may also be generalized to \textit{n}
qubits as $|W_{n}\rangle =a_{1}\left\vert 100...0\right\rangle
+a_{2}\left\vert 010...0\right\rangle +a_{3}\left\vert 001...0\right\rangle
+...++a_{n}\left\vert 000...1\right\rangle $. These states exhibit the
maximal bipartite entanglements and no other \textit{m}-way entanglements,
i.e., 
\begin{eqnarray}
\tau _{m}^{(k)} &=&0\text{ for }2<m\leq n,\hspace{1cm}\tau
_{2}^{(k)}=\sum\limits_{j\neq k}4\left\vert a_{k}a_{j}\right\vert ^{2} 
\notag \\
S_{k}^{2} &=&\left( \left\vert a_{k}\right\vert ^{2}-\sum\limits_{j\neq
k}\left\vert a_{j}\right\vert ^{2}\right) ^{2}
\end{eqnarray}%
In these two cases the complementarity relation generalizes to $%
\sum\limits_{m=2}^{n}\tau _{m}^{(k)}+S_{k}^{2}=1$.

\section{Conclusions}

Complementarity is a universal relationship between properties of quantum
objects. However, it behaves in different ways for different quantum
objects. The purpose of this paper was to analyze the different
complementarity relations that exist in two- and multi-qubit systems and to
illustrate some of them in a simple NMR system.

We experimentally verified the complementarity relation between the
single-particle and bipartite properties: $C^{2}+S_{k}^{2}=1$ in a pure
two-qubit system. To determine the entanglement, we used either the
two-particle visibility $V_{12}$ or the distinguishability $D_{k}$ and the
predictability $P_{k}$. Accordingly, two complementarity relations: $%
V_{12}^{2}+V_{k}^{2}+P_{k}^{2}=1$ and $D_{k}^{2}+V_{k}^{2}=1$ were tested
for different states including maximally entangled, separable, as well as
partially entangled (intermediate) states.

Furthermore, the complementarity $C^{2}+S_{k}^{2}=1$ between one- and
two-particle character was generalized to systems of \textit{n} qubits. The
complementarity relation $C_{k(ij...m)}^{2}+S_{k}^{2}=1$ holds for an
arbitrary pure \textit{n}-qubit state, which implies a tradeoff between the 
\emph{local} single-particle property ($S_{k}^{2}=V_{k}^{2}+P_{k}^{2}$) and
the \emph{nonlocal} bipartite entanglement between the particle and the
remainder of the system ($C_{k(ij...m)}^{2}$). More interesting, in a pure
three-qubit system, the single-particle character ($S_{k}^{2}$), the
two-particle property regarding this particle measured by the sum of all
pair-wise entanglements involving the particle ($\tau _{2}^{(k)}$), and the
three-particle property measured by the \textit{genuine} tripartite
entanglement ($\tau _{3}$) are complementary, i.e., $\tau _{3}+\tau
_{2}^{(k)}+S_{k}^{2}=1$. However, the generalization of the similar
relationship to a larger-qubit system requires the identification and
quantification of multi-partite entanglement for pure and mixed states
beyond three-qubit systems that still remains an open question currently. A
similar relationship cannot be directly generalized to larger qubit systems.
Some specific samples might be helpful to conjecture the relation $%
\sum\limits_{m=2}^{n}\tau _{m}^{(k)}+S_{k}^{2}=1$: the single-particle
property (\emph{local}) of a particle might be complementary to all possible 
\textit{pure} multi-particle properties (\emph{nonlocal}) connected to this
particle.

Complementarity and entanglement are two important phenomena that
characterize quantum mechanics. From these observations, we conclude that
entanglement in its various forms is an important parameter for the
different forms of complementarity relations in multi-partite systems.
Different forms of entanglement quantify the amount of information encoded
in the different quantum correlations of the system, indicating the
multi-partite quantum attributes. These results have also implications on
the connection between entanglement sharing and complementarity and maybe in
turn provide a possible way to study the entanglement in multi-partite
quantum systems by complementarity. We hope that these findings will be
useful for future research into the nature of complementarity and
entanglement.

\begin{center}
\textbf{ACKNOWLEDGMENTS}
\end{center}

We thank Reiner K\"{u}chler for help with the experiments. X. Peng
acknowledges support by the Alexander von Humboldt Foundation. This work is
supported by the National Natural Science Foundation of China (Grant NO.
10274093 and 10425524) and by the National Fundamental Research Program
(2001CB309300).

\end{document}